\documentclass[aip,reprint,graphicx,twocolumn,superscriptaddress,nofootinbib]{revtex4-2}
\usepackage{amssymb,amsmath,amsbsy,graphicx,xcolor,times,subfigure,marvosym,color,bm,multirow,epstopdf,color,soul}

\begin{document}
\title{Non-perturbative cathodoluminescence microscopy of beam-sensitive materials}

\author{Malcolm Bogroff*}
\affiliation{Materials Science and Technology Division, Oak Ridge National Laboratory, 1 Bethel Valley Rd, Oak Ridge, TN 37831}

\author{Gabriel Cowley*}
\affiliation{Materials Science and Technology Division, Oak Ridge National Laboratory, 1 Bethel Valley Rd, Oak Ridge, TN 37831}

\author{Ariel Nicastro*}
\affiliation{Materials Science and Technology Division, Oak Ridge National Laboratory, 1 Bethel Valley Rd, Oak Ridge, TN 37831}

\author{David Levy}
\affiliation{Materials Science and Technology Division, Oak Ridge National Laboratory, 1 Bethel Valley Rd, Oak Ridge, TN 37831}

\author{Yueh-Chun Wu}
\affiliation{Materials Science and Technology Division, Oak Ridge National Laboratory, 1 Bethel Valley Rd, Oak Ridge, TN 37831}

\author{Nannan Mao}
\affiliation{Electrical Engineering and Computer Science Department, Massachusetts Institute of Technology, 77 Massachusetts Ave. Cambridge, MA 02139}

\author{Tilo H. Yang}
\affiliation{Electrical Engineering and Computer Science Department, Massachusetts Institute of Technology, 77 Massachusetts Ave. Cambridge, MA 02139}

\author{Tianyi Zhang}
\affiliation{Electrical Engineering and Computer Science Department, Massachusetts Institute of Technology, 77 Massachusetts Ave. Cambridge, MA 02139}

\author{Jing Kong}
\affiliation{Electrical Engineering and Computer Science Department, Massachusetts Institute of Technology, 77 Massachusetts Ave. Cambridge, MA 02139}

\author{Rama Vasudevan}

\affiliation{Center for Nanophase Materials Sciences, Oak Ridge National Laboratory, 1 Bethel Valley Rd, Oak Ridge, TN 37831}
\author{Kyle P. Kelley}

\affiliation{Center for Nanophase Materials Sciences, Oak Ridge National Laboratory, 1 Bethel Valley Rd, Oak Ridge, TN 37831}
\author{Benjamin J. Lawrie}
\affiliation{Materials Science and Technology Division, Oak Ridge National Laboratory, 1 Bethel Valley Rd, Oak Ridge, TN 37831}
\affiliation{Center for Nanophase Materials Sciences, Oak Ridge National Laboratory, 1 Bethel Valley Rd, Oak Ridge, TN 37831}
\email[]{lawriebj@ornl.gov}

% Affiliations: Please provide adacemic titles (Prof. or Dr.) for all authors where applicable, and include an institutional email address for all corresponding authors
%\begin{affiliations}

\date{\today}

\begin{abstract}Cathodoluminescence microscopy is now a well-established and powerful tool for probing the photonic properties of nanoscale materials, but in many cases, nanophotonic materials are easily damaged by the electron-beam doses necessary to achieve reasonable cathodoluminescence signal-to-noise ratios. Two-dimensional materials have proven particularly susceptible to beam-induced modifications, yielding both obstacles to high spatial-resolution measurement and opportunities for beam-induced patterning of quantum photonic systems.  Here pan-sharpening techniques are applied to cathodoluminescence microscopy in order to address these challenges and experimentally demonstrate the promise of pan-sharpening for minimally-perturbative high-spatial-resolution spectrum imaging of beam-sensitive materials.\footnote{This manuscript has been authored by UT-Battelle, LLC, under contract DE-AC05-00OR22725 with the US Department of Energy (DOE). The US government retains and the publisher, by accepting the article for publication, acknowledges that the US government retains a nonexclusive, paid-up, irrevocable, worldwide license to publish or reproduce the published form of this manuscript, or allow others to do so, for US government purposes. DOE will provide public access to these results of federally sponsored research in accordance with the DOE Public Access Plan (http://energy.gov/downloads/doe-public-access-plan).}
\end{abstract}


\maketitle %\maketitle must follow title, authors, abstract and \pacs

\section{Introduction}
Color centers and localized excitons in two-dimensional (2D) materials have emerged as a promising resource for quantum networking and quantum sensing in recent years~\cite{aharonovich2016solid,atature2018material,liu20192d,turunen2022quantum} because of the potential for atomic scale control over the defect environment~\cite{dyck2018building,dyck2023atom}, compatibility with integrated photonic circuits\cite{turunen2022quantum,pelucchi2022potential,liu20192d,spencer2023monolithic}, and the potential for manipulation of emitter photophysics with engineered strain environments~\cite{luo2023deterministic,luo2023imaging,parto2021defect,iff2019strain,curie2022correlative} and electrical gating~\cite{white2022electrical,luo2023improving,schadler2019electrical}. Unfortunately, many reports in the literature focus on `hero' emitters that are selected after exhaustive searches of many lower quality emitters. Attempts to locate and pattern individual single photon emitters with desirable brightness, purity, and indistinguishability often result in the observation of multiple emitters within a single diffraction-limited spot or in the emergence of coupled electronic transitions with unwanted photochromic effects~\cite{feldman2019phonon,feldman2020evidence}. The ability to manipulate and measure the quantum states associated with these color centers relies heavily on our understanding of how nanoscale heterogeneities affect their photophysical behavior. Therefore, advanced nanoscale probes that can accurately assess these effects while allowing for in situ modification are crucial to the development of color centers for practical quantum technologies.

\begin{figure*}[hbt]
  \includegraphics[width=\linewidth]{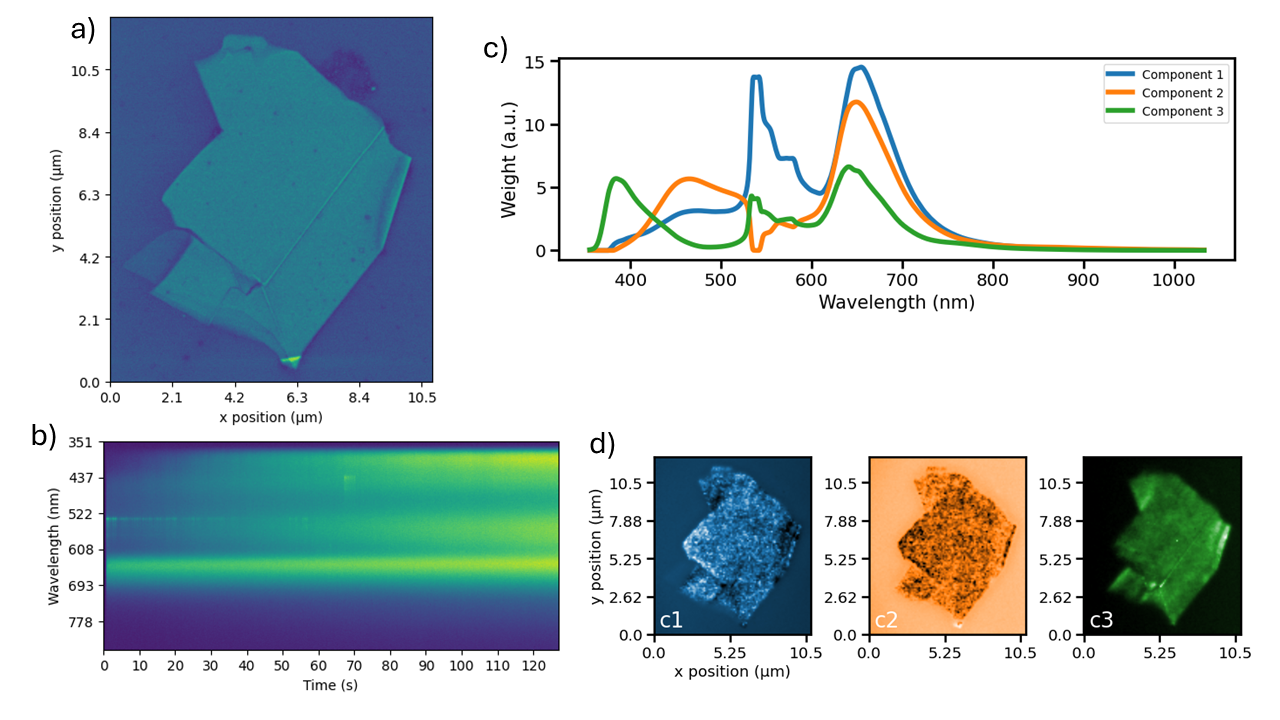}
  \caption{(a) SEM image of an hBN flake with horizontal width of 10.5 microns. (b) Time-series CL spectra acquired at a single point on the hBN flake highlighting the beam-induced changes in the hBN CL spectrum as a function of increasing dose. (c) Spectral components and (d) intensity maps generated by non-negative matrix factorization of CL spectrum image acquired in conventional rastered CL spectrum imaging modality.}
  \label{fig:fig1}
\end{figure*}

Cathodoluminescence (CL) microscopies have emerged as a powerful nanoscale probe of quantum nanophotonic systems \cite{luo2023imaging,curie2022correlative,bonnet2024cathodoluminescence,roux2022cathodoluminescence,zheng2017giant,fiedler2021brobdingnagian,iyer2021near}. The converged electron-beam probe offers a nanometer-scale excitation, and far-field collection of CL enables high sensitivity measurements of emitter energetics and dynamics across a wide variety of energy- and time-scales. However, the electron-beam probe has also emerged as a resource for beam-induced modification of 2D materials \cite{roux2022cathodoluminescence,
horder2022coherence,gale2022site,bianco2023engineering}. Indeed, many monolayer transition metal dichalcogenides only exhibit measurable CL signals when they are encapsulated by hBN \cite{zheng2017giant,fiedler2021brobdingnagian}, an effect that may result from beam-induced damage to beam-sensitive materials. While clear examples exist in the literature using the electron beam to either probe or manipulate color centers in 2D materials, the necessary electron-beam dose to measure emitter photophysics is in many cases also sufficient to substantially modify the defect environment in that material. Thus, identifying new minimally-perturbative approaches to CL microscopy capable of probing the color center environment without modifying it --- while allowing for intentional in situ modification at higher electron-beam doses --- is critical to improved understanding and control of 2D quantum photonic systems.

%Most images are highly compressible because they can be represented sparsely in some basis. While images are frequently compressed after acquisition to reduce file size, compressive sensing (CS) schemes rely on van appropriate selection of sampling matrices and reconstruction algorithms in order to allow for accurate image reconstruction with many fewer measurements than are required by the Shannon-Nyquist sampling theorem\cite{willett2011compressed,rani2018systematic}. CS schemes have been used to enable new imaging modalities in astronomy\cite{bobin2008compressed}, for new types of quantum imaging and quantum process tomography \cite{jiying2010high,shabani2011efficient,lawrie2013toward}, and for scanning probe and scanning tunneling microscopies\cite{lerner2021compressed,kelley2020fast}. The most well-known CS schemes rely on a `single-pixel camera' approach wherein a spatial light modulator or similar element is used to sample random array of pixels from an image before the integrated intensity from those pixels is measured on a single pixel detector. This approach is repeated \textit{N} times before the original image is reconstructed from the composite set of sampling matrices and measured intensities. However, other approaches relying on the same basic premise of image compressibility may acquire data at each pixel of a sampling pattern\cite{lerner2021compressed,kelley2020fast}.

Pan-sharpening (PS) methods may offer a minimally perturbative approach to CL microscopy by combining separate high-spatial-resolution and high-spectral-resolution images in order to generate a composite image with both high spatial and spectral resolution. First used in satellite imaging~\cite{javan2021review,loncan2015hyperspectral}, PS methods have now emerged as powerful tools for multidimensional imaging in a wide variety of use cases. In nanoscience, PS has been applied to electron energy loss spectroscopy~\cite{borodinov2021eels}, scanning probe microscopy~\cite{borodinov2019afmir}, and secondary ion mass spectrometry~\cite{tarolli2014application}, though PS-CL has not yet been explored despite the substantial benefit associated with minimizing beam-induced damage through undersampling of hyperspectral CL images. Most PS algorithms rely on either (i) the substitution of spectral components from a hyperspectral dataset with a high spatial resolution panchromatic image or (ii) a multiresolution analysis approach based on injection of spatial details from the panchromatic image into resampled hyperspectral bands~\cite{vivone2014critical}. Here, we focus on PS-CL performed using the Brovey transform~\cite{gillespie1987color}, an example of the former class of PS algorithms that uses multiplicative sharpening to spatially modulate spectral pixels \cite{vivone2014critical}, and we examine the impact of this approach on CL imaging of color centers in 2D materials.

\section{Methods}

Hexagonal boron nitride (hBN) is known to be relatively robust to electron-beam exposure, and it has been probed by conventional CL microscopy \cite{curie2022correlative}, but there is also a growing literature describing electron-beam induced patterning of color centers in hBN \cite{roux2022cathodoluminescence,
horder2022coherence,gale2022site,bianco2023engineering}. Thus, hBN offers a valuable platform for examining PS-CL techniques that could be crucial to probes of more environmentally sensitive materials like monolayer transition metal dichalcogenides and some classes of hybrid organic perovskite thin films. All data reported here is based on exfoliated hexagonal boron nitride (hBN) flakes transferred onto a 300 nm silicon dioxide (SiO\textsubscript{2}) layer on a silicon substrate.

Cathodoluminescence data was acquired using a Delmic Sparc CL module with an FEI Quattro scanning electron microscope (SEM) operating with a beam energy of 5 kV and a beam current of 110 pA at room temperature and a chamber pressure of 1E-6 Torr. CL spectrum images were acquired with an Andor Kymera spectrograph and an Andor Newton CCD with an acquisition time of 300 ms per spectrum. A pickoff mirror was used to direct the collected CL signal into a photomultiplier tube (PMT) for high-spatial resolution panchromatic CL imaging using a PMT integration time of 10 $\mathrm{\mu}$s (yielding 30,000x reduced dose per pixel compared with spectrum imaging).

\begin{figure*}[hbt]
  \includegraphics[width=\linewidth]{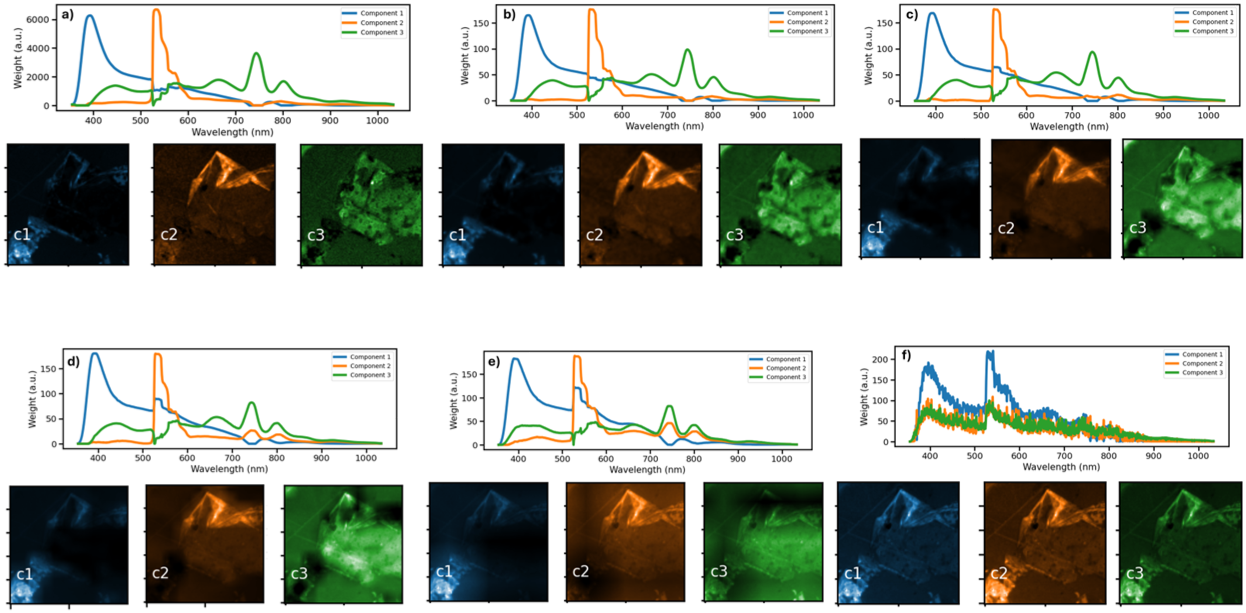}
  \caption{Benchmark PS-CL results generated from a single CL hyperspectral image of an hBN flake with dimensions of 100x100x1024 (horizontal, vertical, wavelength) pixels with a horizontal field of view of 30$\mu$m. A panchromatic image was generated from the spectrum image by summing along the wavelength axis while spectrum images with reduced spatial resolution were generated by binning spatial pixels together. The Brovey pan-sharpening algorithm was used to reconstruct a hyperspectral CL image from these datasets. NMF reconstructions of the pan-sharpened CL images are shown for data generated from the panchromatic image and (a) the complete 100x100x1024 spectrum image, (b) a 50x50x1024 spectrum image, (c) a 25x25x1024 spectrum image, (d) a 12x12x1024 spectrum image, (e) a 6x6x1024 spectrum image, and (f) a 1x1x1024 spectrum image.}
  \label{fig:fig2}
\end{figure*}

\section{Results}

An SEM image of a prototypical hBN flake is illustrated in Fig.~\ref{fig:fig1}(a). While we were able to acquire moderately coarse spatial resolution CL spectrum images of hBN flakes with no apparent degradation (with pixel sizes of order 100 nm), improving spatial resolution while maintaining the beam energy and current along with a constant dwell time per pixel resulted in growing evidence of beam-induced modification of the hBN flake. This dose-dependent beam-induced modification is most easily visualized through time-series spectra acquired while the electron beam rapidly scanned 768x512 pixels across a 500 nm spot with a 100 ns/pixel electron dwell time. Note that acquiring time-series spectra with the electron beam focused on a single spot resulted in immeasurably fast changes in the CL spectra, so averaging across a 500 nm spot allowed us to reduce the effective dose/pixel during time-series CL spectrum acquisition. A reduced CL spectrum acquisition time of 100 ms (approximately 2.5x the time required to scan 768x512 pixels) was used in order to monitor the time-dependent changes in CL spectra as a function of electron-beam dose.

Several features are immediately apparent in the time-series spectra shown in Fig.~\ref{fig:fig1}(b): A prominent CL band centered at a wavelength of 660 nm exhibits minimal change with increasing dose. On the other hand, a blue CL band near 408 nm grows monotonically with increasing dose, and the CL from a narrow linewidth color center near 548 nm disappears with increasing dose. These results highlight the importance of alternative SEM-CL acquisition modalities, especially for smaller pixel sizes where the increased electron-beam dose can result in rapid modification of the hBN color center photophysics.  Additionally, it is difficult to interpret CL spectra acquired at a single point, as we expect some CL contribution from defect bands in the SiO\textsubscript{2} substrate.

\begin{figure}[hb!]
  \includegraphics[width=0.9\columnwidth]{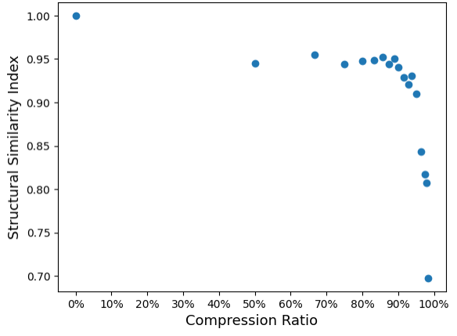}
  \caption{Calculated structural similarity index as a function of compression ratio for the PS-CL data shown in Fig. 2. The compression ratio is calculated based on the compression of the hyperspectral data prior to pan sharpening.}
  \label{fig:fig3}
\end{figure}

\begin{figure*}[hbt]
  \includegraphics[width=0.9\linewidth]{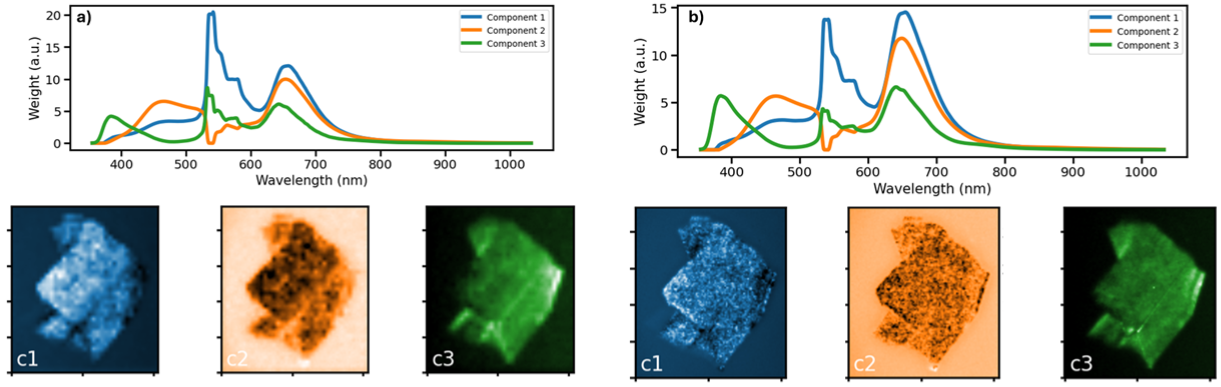}
  \caption{Three component NMF reconstructions of hBN PS-CL images generated from (a) an 1168$\times$1034 pixel panchromatic image and a 39$\times$35 pixel hyperspectral image and (b) an 1168$\times$1034 panchromatic image and a 117$\times$105 pixel hyperspectral image. The horizontal field of view is 10.5 $\mu$m.
  }
  \label{fig:fig4}
\end{figure*}

Figures~\ref{fig:fig1}(c) and (d) illustrate a non-negative matrix factorization (NMF) decomposition of a CL spectrum image acquired across this flake using conventional raster scanning with a pixel size of 100 nm. For all NMF decompositions shown in this manuscript, reconstructions were attempted with varying numbers of components, and it was determined that three components were sufficient to provide a reasonable reconstruction of the raw data. The NMF decomposition shown in Fig.~\ref{fig:fig1} immediately aids in the interpretation of the single point CL spectra shown in Fig.~\ref{fig:fig1}(b): Component 3 is solely a result of the hBN flake, while Component 2 appears to be primarily a result of the SiO\textsubscript{2} substrate luminescence. Component 1, which features the narrow transition at 548 nm, appears to be primarily due to the hBN flake, though it includes some convolution with substrate luminescence.  The narrow band color centers seen in Component 1 are reasonably densely distributed across the hBN flake.  Unfortunately, it is hard to image these color centers with improved spatial resolution using conventional CL raster scanning because reducing pixel sizes while maintaining the field of view yields a combination of beam-induced damage and unacceptably long measurement times. However, PS-CL techniques offer a promising pathway to address this challenge.

The Brovey transform was identified as a PS algorithm well suited to our data, and it was benchmarked by generating high spatial resolution and high spectral resolution datasets from a single 100$\times$100$\times$1024 (horizontal $\times$ vertical $\times$ wavelength) pixel hyperspectral CL image of a hBN flake (by separately binning all wavelengths together to create a 100$\times$100 pixel panchromatic image and binning adjacent spatial pixels to create spectrum images with 1024 spectral pixels and between 1$\times$1 and 50$\times$50 spatial pixels). We then up-sampled the hyperspectral dataset to match its spatial resolution with that of the panchromatic image using the resize function in skimage.transform with linear splines. Each spatial pixel's spectrum in this new dataset is scaled to match the net intensity of the corresponding normalized panchromatic pixel's intensity. The pan-sharpened CL spectrum image then had dimensions of 100$\times$100$\times$1024 and could be easily compared with the original spectrum image.

Three component NMF reconstructions of the PS-CL image with reduced dimensionalities are shown in Fig. \ref{fig:fig2}. At first glance, very little information is lost in the PS-CL image as the hyperspectral image is collapsed from a 100$\times$100$\times$1024 spectrum image (Fig. \ref{fig:fig2} (a)) to a 6$\times$6$\times$1024 image (Fig. \ref{fig:fig2} (e)), though unsurprisingly, all three components of the PS-CL image generated from a 1$\times$1$\times$1024 spectrum image (Fig. \ref{fig:fig2} (f)) look nearly identical to one another.  The quality of the pan-sharpening algorithm can be calculated here with a structural similarity index (SSI) comparing the pan-sharpened image with the original 100$\times$100$\times$1024 spectrum image. The SSI of each pan-sharpened image here was calculated using the scikit-image library. As shown in Fig.~\ref{fig:fig3}, applying the Brovey transform to the original 100$\times$100$\times$1024 spectrum image results in a SSI of 1.0, and a SSI$>0.9$ for compression ratios as high as 90\%.

This baseline PS-CL data suggests that CL spectrum images can be acquired with substantially reduced electron-beam exposure by combining short-dwell-time panchromatic CL images acquired on a PMT with very low spatial resolution spectrum images. Further, rastering the electron-beam over each spectrum-image pixel during the comparably-slow spectrum acquisition time will distribute the electron-beam dose over a large area and minimize the risk of beam-induced damage. 

With this understanding in hand, PS-CL images were reconstructed from raw hyperspectral and panchromatic CL datasets. Figure~\ref{fig:fig4} illustrates three-component NMF reconstructions of PS-CL images of an hBN flake generated from an 1168$\times$1034 panchromatic CL image (using 10 $\mu$s dwell time per pixel) and a 39$\times$35 pixel spectrum image (Fig.~\ref{fig:fig4}(a)) and a 117$\times$105 pixel spectrum image (Fig.~\ref{fig:fig4}(b)) (each using a 300 ms spectrum acquisition time per pixel).  Both exhibit very similar spectral components, though the latter does exhibit higher-spatial-resolution intensity maps. Nonetheless, these results show the potential impact of pan-sharpening for CL microscopy with minimal beam-induced damage.

\section{Conclusion}
The PS-CL results shown here suggest that hyperspectral CL measurements can be undersampled by 90\% while maintaining at least 90\% fidelity to the ground truth. Because hyperspectral and panchromatic datasets can be easily acquired concurrently (using a beamsplitter) or consecutively (using a pickoff mirror) with no change to the experimental alignment, the approach described here is easily adapted for a wide variety of CL microscopies of beam-sensitive materials. This approach is particularly critical for minimizing beam-induced damage in beam-sensitive materials like monolayer 2D materials and hybrid organic perovskite thin films, and for applications that require high spatial resolution and long spectrum acquisition times. The successful demonstration of PS-CL also unlocks new opportunities for high-dose electron-beam patterning of single photon emitters with low dose PS-CL in situ characterization.

\medskip
%\textbf{Supporting Information} \par %Please delete the Suppporting Information statement if it is not applicable. Please supply Supporting Information in another file. Supporting information should not be provided in .tex format
%Supporting Information is available from the Wiley Online Library or from the author.

% Acknowledgements
\begin{acknowledgments}
This research was supported by the U. S. Department of Energy, Office of Science, Basic Energy Sciences, Materials Sciences and Engineering Division. CL microscopy and data analytics algorithms were supported by the Center for Nanophase Materials Sciences, which is a U.S. Department of Energy Office of Science User Facility. MB, GC, AN, and DL were supported by the U.S. Department of Energy, Office of Science, Office of Workforce Development for Teachers and Scientists (WDTS) under the Science Undergraduate Laboratory Internship program. 
\end{acknowledgments}

% References

% Use the following code if you wish to generate your bibliography with BibTeX;
% replace the string "MSP-template" below with the name(s) of
% the BibTeX data base(s) you want to use.
% The resulting bibliography-output (the content of the .bbl file)
% must be pasted back into this file before submission.
% Please also include your BibTeX data base file(s) in your submission
% so that we can re-run BibTeX if necessary.
%
%\bibliography{references}
%aipnum4-2.bst 2019-01-14 (MD) hand-edited version of apsrev4-1.bst

% Table of contents entry should be 50 - 60 words long
% Image should be 55 mm broad and 50 mm high or 110 mm broad and 20 mm high

\end{document}